**Directed destabilization of lysozyme in protic ionic liquids reveals a compact, low energy, soluble, reversibly-unfolding (pre-fibril) state.**

Nolene Byrne, Jean-Philippe Belieres and C. Austen Angell

Dept of Chemistry and Biochemistry, Arizona State University, Tempe AZ 85287, USA.
Corresponding author caa@asu.edu

**Recent demonstrations of extraordinary stabilization of proteins in mobile protic [1] and aprotic [2] ionic liquid solutions at ambient temperatures have raised hopes of new biopreservation and drug transportation technologies. Here we examine the relation of folded protein stability to the state of the transferred proton [1], as determined by the N-H proton chemical shift, δ(N-H). We identify a range of δ(N-H) in which the unfolded lysozyme refolds 97%. Exceeding the stability range in the "acid" direction leads to the sudden formation and stabilization of a small, soluble, amyloid form of lysozyme which has its own stability range and which can again unfold/refold many times before an irreversible process, fibrillization, occurs. The tightly bound amyloid form of the lysozyme molecule, identified by circular dichroism spectra and dynamic light scattering, must be of very low energy since the unfolding process absorbs almost three times the enthalpy of "normal" lysozyme unfolding. α- lactalbumin shows similar behavior.**

Protic ionic liquids (PILs) form a subgroup of the "ionic liquid" family that is currently the focus of much research activity. Whereas aprotic ionic liquids have, in common, low vapor pressures and ionic conductivities that are proportional to their fluidities, their protic cousins can vary enormously in properties depending on the proton transfer free energy involved in their formation. The proton transfer energy can be roughly predicted from aqueous $pK_a$ data [3], and, in the case of PILs formed by proton transfer to a nitrogen base site (linear or cyclic amines), can be directly indicated by the N-H proton chemical shift, $\delta(N-H)$, which depends on the nitrogen-derived electron density sensed by the attached proton. The smaller the transfer energy the less firmly the proton is attached to the nitrogen and so the more upfield from an internal TMS standard the $\delta(N-H)$ will lie. (At the limit of weak transfers, the strength of the resonance will die off as the protons distribute themselves between the nitrogen and the original site on the acid molecule).

It is implied that, notwithstanding the large (2-10 M) concentration of protonated species in PILs, the chemical activity of the proton can vary enormously, according to the transfer energy. A typical moist litmus paper will register red in triethyl ammonium triflate, but dark blue in triethyl ammonium acetate. Electrochemical methods of measuring the associated proton activity (a PIL equivalent of the pH of aqueous solution) are under active development [4] but, as these remain subject to change as the optimum calibration methods are determined, we will here use the $\delta(N-H)$ as an unambiguous indicator of the state of the protons in our solutions. Our ability to tune the proton activity (PA), by variation of the PIL or of the composition of binary pIL solutions, is crucial to the present study.

In recent work [1] we showed that, in PILs of comparable PA (i.e. similar N-H chemical shifts [4],), proteins such as lysozyme and ribonuclease A could be unfolded and refolded many times with very little loss of protein to irreversible aggregation. Here we show, by creating solutions of variable PAs (quantified by δ(N-H)), that there is a specific range of PAs that stabilize each protein. Outside this range both the unfolding enthalpy, and the "refoldability" which we now define, drop off rapidly.

For our present purposes, the stability of the protein needs to be assessed without a waiting period of months or years [1]. Following ref 1, we will assume that the stability can be quickly assessed in terms of the refolding fraction, i.e. the fraction of the initial amount of protein that is observed by calorimetry to be unfolding after a prior thermal unfolding - as described in ref 1. For the hen white lysozyme (HWL) that we have studied most, this refolding fraction (stability index) can be as large as 0.97 per cycle, and we have seen comparable values in the cases of ribonuclease A and α-lactalbumin.

Our most interesting finding concerns observations made during controlled excursions beyond the high stability range, where we find that a new form of the protein, of much lower energy, can be generated. This form, which we will characterize, also proves capable of reversible unfolding with a high refolding fraction. However, on a much longer time scale, it generates fibrils. This behavior, which is related to that observed in acidified aqueous systems [5] but differs critically in the lifetime of the molecular intermediate, is not unique to lysozyme but is also seen with its similarly sequenced protein cousin, α- lactalbumin.

In Figure 1 we show the variation of the folding fraction for lysozyme with change of δ(N-H) as the composition of the ionic liquid component is varied. The filled symbols represent observations in different pILs, identified by numbers, each containing a fixed 20 wt% water content. The open symbols represent solvent compositions of the same water content within two binary mixtures (1) EAN (weakly acid) –TEATf (more acidic) and (2) EAN – EAFm, (mildly basic). Together, these cover a wide range of PAs. One notes a rather broad zone of δ(N-H) in which the refolding fraction is roughly constant at 0.97, with rapid fall-off on either side. It is clear that this high refolding fraction does not depend on the presence of the "folding aid" EAN [6] but rather is obtained with any PIL, pure or solution, that provides a PA in the stable range. Similar results have been obtained when the water content is varied, changing the δ(N-H). Again a refolding fraction maximum appears for the same δ(N-H) range. Thus, the PA rather than the actual water content or the actual ionic liquid composition, is identified as the important variable – which causes no real surprise. However solutions containing a high water content are not expected to show the long term stability reported in ref. 1, due to hydrolysis.

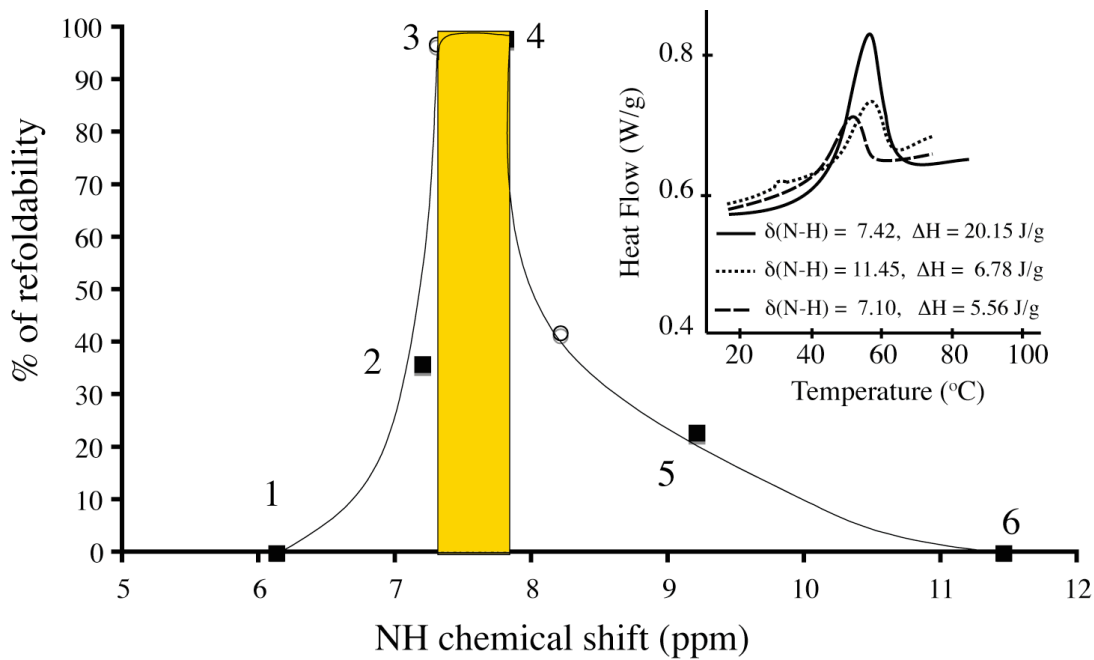

**Figure 1.** Refolding fraction (defined by the ratio of area under the DSC unfolding endotherms of second to first upscans after change of PA represented by the proton chemical shift δ(N-H) referenced to external tetramethyl silane. **Insert.** Unfolding endotherms in the most stable domain compared with those on either side (δ(N-H) values of 7.42 and 11.45). 1: EAHSO$_4$, 2: TEATf; 3: EAN; 4: HOEAN, 5: EAFm and 6: DBAAc. Open circles are 1:1 binary mixtures of EAN-TEATf, EAN-TEATfAc and EAN-EAFm

It is interesting to observe the effect on the **enthalpy** of the refolding process as the refolding fraction is reduced by change of PA. As the high stability zone in Figure 2 is exited, in either direction, the enthalpy of unfolding strongly decreases. Unlike the refolding fraction, however, it does not approach zero but rather levels off near the smaller values seen in the Figure 2 insert. This is similar to reports on the effect of pH on denaturation enthalpies (and "melting" temperatures) of lysozyme and other proteins in familiar aqueous solution systems[7,8].

On changing the PA of the solution to strongly acid values, (by using ammonium hydrogen sulfate as the ionic liquid, at water content 20 wt%), an unexpected observation is made during the upscan. Instead of the apparent heat capacity starting to increase with onset of unfolding as in Fig. 1 insert, it instead strongly decreases to a lower energy state as shown in the main part of Figure 2 (curve 1). The exotherm, which occurs over the temperature range 60 °C -95 °C amounts to 65 J/g lysozyme, i.e. nearly three times the enthalpy of unfolding in the most stable region of the normal protein (seen in Figure 1 insert). The final state of this initial scan in Fig. 2 is not a truly stable state of the system since it continues to relax slowly to lower energies over the next few heat/cool cycles. It then stabilizes so as to give an upscan similar in form to that seen in the normal unfold/refold zone of Figure 1 but with total energy absorbed now larger by almost a factor of ~3 (57.4 J/g for new low energy state vs. 20.2 J/g for normal HWL - as judged by Fig.1 insert). α-lactalbumin, which has a primary sequence closely related to that of lysozyme, shows related behavior, also exhibiting the large enthalpy unfolding endotherm but after less initial cycling. This implies that the refolding process is partly achieved in the initial ambient temperature dissolution process.

This new unfolding endotherm is highly reversible, as illustrated by the three successive heat/cool cycles in the insert of Figure 2. The unfolding entropy in this solution must also be high, in order to understand that the unfolding occurs with little change in endotherm peak temperature. Indeed, the peak temperature is almost the same as for lysozyme in biological conditions.

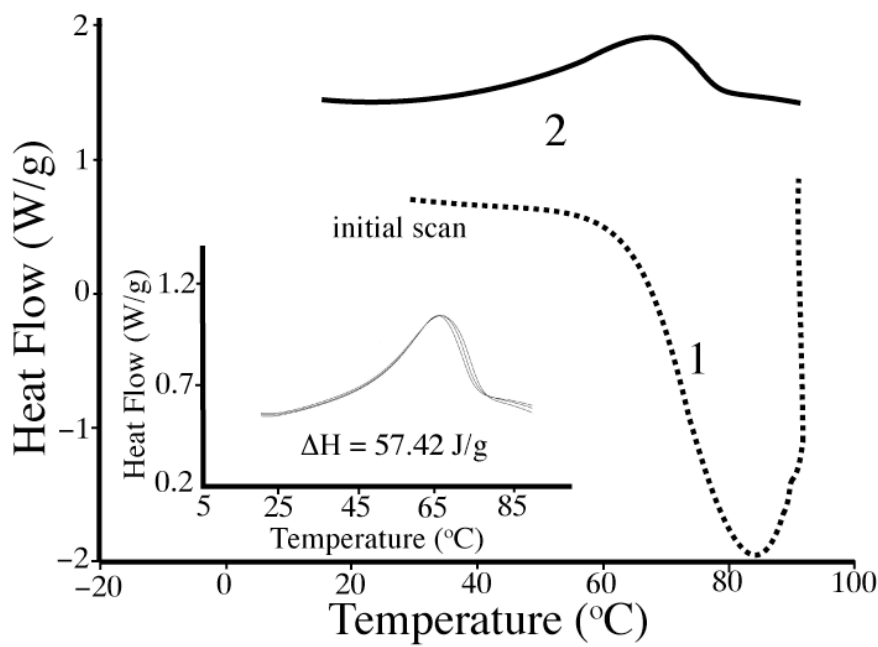

**Figure 2.** Apparent $C_p$ for upscan of lysozyme in $NH_4HSO_4$ + 20wt% $H_2O$ for (1) initial scan showing the large exotherm and (2) final 'low energy' state which has been reached after several heat/cool scans. **Insert** shows three successive unfolds of the new low energy state with very little loss per cycle.

It is evident that some new and rather stable state of lysozyme has been obtained by the above procedure. It is of interest to know what its structure might be. Accordingly, a study of the near-UV circular dichroism spectra, which are often used to assess the proportions of α-helix vs β-sheet structure of proteins[9-11], has been made. Results for the new low energy state before and after the unfolding endotherm of Figure 2 insert, are shown in Figure 3 as thick solid and thin solid lines, respectively. The minimum at 215 cm$^{-1}$, which disappears during the endotherm, is typical of the near UV CD spectrum for fibrillized HWL reported by Daoying Hu, and Luhua [12]. The same minimum, 215 nm, has been reported for fibrilized myoglobin by Fändrich et al. [9] The 215 nm minimum in the CD spectrum is regarded as the prime indicator of β-sheet structural elements, so our low energy form of lysozyme is evidently rich in β-sheet structure but, remarkably, has remained capable of reversible unfolding.

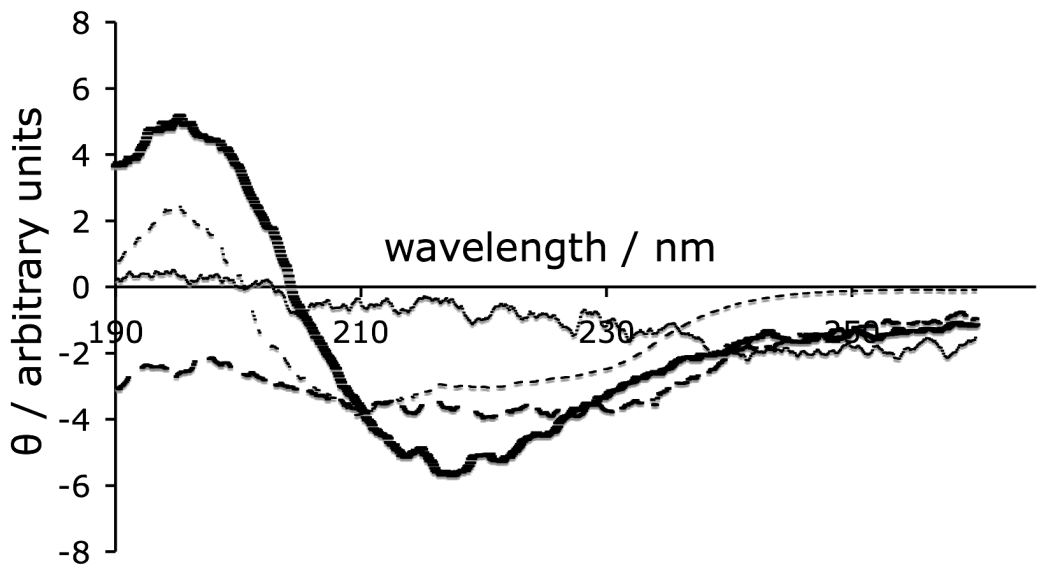

**Figure 3.** Near-UV Circular Dichroism spectra for lysozyme dissolved in $NH_4HSO_4$ (1) the initial state before heating scan 1 of Fig. 2 (thick dashed line), (2) stabilized low energy configuration before the unfolding scan, showing β-sheet minimum at 215nm (thick solid line) and (3) after unfolding (thin solid line). The thin dashed line is the spectrum of folded HWL in aqueous solution. (215 $nm^{-1}$ is also the minimum for high ethanol content, fully fibrillized, HWL).

A critically important question arises. Has the ionic liquid medium stabilized the much-discussed [13-15] amyloid molecular precursor to fibrillization? The likelihood of this being the case is supported by the observation that, although unchanged over a two week period at ambient temperature, the strength of the unfolding isotherm of Figure 2 insert diminishes over a longer period and vanishes after a month. This is presumably due to fibrillization, but the necessary electron microscope characterization has not yet been carried out to prove this. If the solution is maintained at 80ºC the fibrillization is completed within hours, with the solution turning white.

If the above scenario is correct, the entity which exhibits the large unfolding enthalpy should be small as well as soluble. The existence of a small precursor to fibrillization has already been clearly indicated by the accelerating effect of pressure on fibrillization kinetics reported by Ferrao-Gonzalez et al [16]. To answer this question for our case, we turn to a simple dynamic light scattering technique that gives precise information on the effective size, or size distribution, of dissolved or suspended macromolecules when the viscosity of the medium is known. In the absence of viscosity data we can still determine relative size distributions. In Figure 4 insert we show data for a solution of lysozyme in EAN 20wt% $H_2O$ solution for which we have an approximate viscosity from ref. 1.

In the main part of Figure 4 we show data for the lysozyme in $NH_4HSO_4$ in three conditions. The first is the initial unfolded state (solid circles), the second (closed squares) is after cycling to produce the low energy amyloid form (the one that can be reversibly unfolded by heating, as seen in Fig. 2 insert). The third state (open squares) is the high temperature unfolded form. For the lower panel samples, the viscosity is much higher but has not been measured. Here we have kept the same viscosity input parameter for the data treatment as for the EAN solution, knowing that this will make the molecule appear smaller by about an order of magnitude. The unfolded forms should be of comparable size. The important observation is that the folded amyloid form is small and narrowly distributed relative to the unfolded forms and even relative to the normal α-helix rich form of lysozyme.

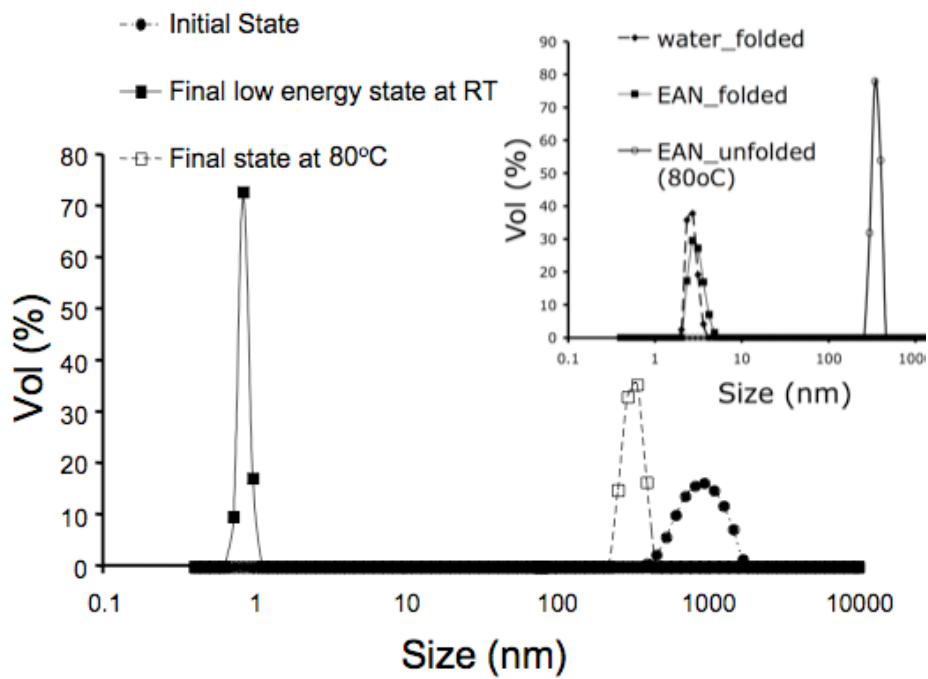

**Figure 4.** Dynamic light scattering based estimates of the volume distribution of particles in the protein solutions in folded and unfolded states. **Main part:** (closed dot) Initial unfolded state in liquid $NH_4HSO_4$ at ambient temperature: (closed square) the refolded β-sheet-rich form at ambient temperature: (open square) unfolded amyloid form at 80°C. **Insert: (**small solid circle**)** folded lysozyme in normal physiological solution: (solid square) folded lysozyme in EAN 20wt%$H_2O$ (open circle) unfolded lysozyme in EAN 20wt%$H_2O$ at 80°C. The particle sizes registered are smaller than the actual by the ratio of the $NH_4HSO_4$ solution viscosity (unknown but larger than for EAN) to that of the EAN solution. Note the narrow distribution for the folded amyloid form. It is smaller relative to its unfolded state and also relative to the normal (α-helix rich) folded form, but we cannot make an absolute comparison of the two folded forms at this time.

It would of course be interesting to know about the stability of this new state of lysozyme in biological solutions. However, we have postponed such a study pending toxicology evaluations on the basis of our reading that the soluble forms of amyloid proteins may be the infectious agents in folding diseases [17].

Finally we wish to note that the fibrillization process we have observed shows an interesting parallel to the process of crystallization reported recently for some other complex systems, liquid germanium, silicon, and water. In each case the process involves formation of an intermediate polyamorphic form which is closer in structure to the crystal than is the form stable at higher temperature. Both are examples of the venerable Ostwald step rule, which states that the path to a final product often passes through a series of intermediates, each of lower free energy than the initial state but kinetically more accessible than the final state. The soluble amyloid form, then, is the first Ostwald step on the route to the lowest free energy state for the lysozyme molecule, the fibril.